\documentclass[11pt]{article}
\usepackage{moriond,epsfig}

\def\be{\begin{equation}}
\def\ee{\end{equation}}
\def\bea{\begin{eqnarray}}
\def\eea{\end{eqnarray}}

\def \D0 {D\O }

\begin{document}
\vspace*{2.7cm}
\title{Higgs Searches at the Tevatron
\footnote{Presented at Recontres de Moriond EW 2007, 10-17 March 2007, La Thuile.}
}

\author{Krisztian Peters\\(on behalf of the CDF and \D0 \ Collaborations)}

\address{School of Physics and Astronomy, The University of Manchester,\\
Oxford Road, Manchester M13 9PL, United Kingdom}

\maketitle\abstracts{Recent preliminary results obtained by the CDF and \D0 \ Collaborations on
searches for the Higgs boson in $p\bar p$ collisions at $\sqrt s =
1.96$ TeV at the Fermilab Tevatron collider are discussed. The data,
corresponding to integrated luminosities of about 1 fb$^{-1}$, show no
excess of a signal above the expected background in any of the decay
channels examined. Instead, upper limits at 95\% Confidence Level on
the cross section are established. Further, a combined Standard Model
Tevatron cross section limit is presented.}

\section{Introduction}

The Higgs boson is the last missing particle in the Standard Model
(SM). Its mass is not determined by the SM, there are however several
experimental constraints which bound the Higgs mass to values which
are within the reach of the Tevatron collider. Lower bounds are given
from direct searches at LEP2. These results exclude Higgs masses below
114.4 GeV at the 95\% Confidence Level (C.L.) \cite{lep}. An upper bound
on the Higgs mass is obtained by global electroweak fits. Especially
radiative corrections to the $W$ mass from the Higgs and top quark
play an important role. New precision measurements of the $W$ mass
\cite{top} and the top mass \cite{W} from the Tevatron favor a light SM
Higgs boson and yield an upper value of 144 GeV at 95\% C.L. (or 182
GeV if the LEP2 limit is included) \cite{ewfit}.

\section{Experimental environment}

The Higgs searches are crucially dependent on performance of the
Tevatron accelerator and detectors. Both, CDF and \D0 \ detectors are
currently performing close to their optimal design values, taking data
with an efficiency of about 90\%. The present Tevatron performance is
matching the design values in terms of the current weekly integrated
and peak luminosity. As of today, more than 2.5 fb$^{-1}$ have been
delivered, with weekly integrated luminosity routinely reaching 50
pb$^{-1}$. If the accelerator keeps following the designed luminosity
evolution, an integrated luminosity of about 8 fb$^{-1}$ will be
achieved by the end of 2009, increasing the potential for a Higgs
discovery at the Tevatron significantly.

\section{Standard Model Higgs searches}

Production cross sections for the SM Higgs boson at the Tevatron are
rather small. They depend on the Higgs mass and are about 0.1 -- 1 pb
in the mass range of 100 -- 200 GeV. The largest production cross
section comes from gluon fusion, where the Higgs is produced via a
quark loop. The second largest cross section, almost an order of
magnitude smaller, is the associated production with vector bosons.
At the mass range covered by the Tevatron, below 135 GeV the highest
branching ratio is given by the decay to $b\bar b$ pairs and for
masses above 135 GeV the Higgs boson decays mainly to $WW$ pairs.

These production and decay properties lead to the following search
strategy at the Tevatron:
\begin{itemize}

\item For masses below 135 GeV the main search channels are the
associated productions with vector bosons where the Higgs decays into
$b\bar b$ pairs. In order to isolate the main background processes to
these channels, an efficient b-tagging algorithm and a good dijet mass
resolution are essential. The same final state produced via the gluon
fusion process leads to a higher cross section but is overwhelmed by
the huge multijet QCD background at a hadron collider.

\item For masses above 135 GeV the search is mainly focused on the gluon
fusion production process where the Higgs decays into $WW$ pairs.

\end{itemize}

\subsection{$WH\to \ell\nu b\bar b$, $\ell=e,\mu$}

For SM Higgs searches the most sensitive production channel at the
Tevatron for a Higgs mass below 135 GeV is the associated production
of a Higgs boson with a $W$ boson. Dominant backgrounds to the $WH$
signal are $W$ + heavy flavor production, $t\bar t$ and single-top
quark production. Both, CDF and \D0 \ performed cut based analyses
with a rather similar approach. Both, electron and muon channels are
studied here. The channels are separated in events having exactly one
"tight" b-tagged jet (ST), and those having two "loose" b-tagged jets
(DT) (with no overlap). The resulting four channels are analyzed
independently to optimize the sensitivity and are later combined. Both
experiments select events with isolated electrons or muons with
$p_T>20$ GeV, require missing transverse energy above 20 GeV and two
jets with $p_T>20$ GeV (\D0 ) or $p_T>15$ GeV (CDF). Cross section
limits are derived from the invariant dijet mass distribution of the
four individual analyses of each experiment and later combined. For
$m_H=115$ GeV the observed (expected) limit is 1.3 (1.1) pb at \D0 \
\cite{WHcutD0} and 3.4 (2.2) pb at CDF
\cite{WHcutCDF}, to be compared to the Standard Model cross section
expectation of 0.13 pb. Thus the best expected measurement is a
factor 8.8 higher than the SM expectation.

\D0 \  analyzed this channel also with the Matrix Element
technique to separate signal from background. Like in the cut based
analysis the four channels ($e$,$\mu$,ST,DT) are analyzed separately
and later combined. The matrix-element-based technique attempts to
make use of all the available kinematic information in the event to
separate signal and background. Therefore leading order Matrix Elements
are used to compute the event probabilities for signal and
background. The present selection criteria is based on the single top
search \cite{} and will be optimized in the future. Although this
selection is not optimal for $WH$, the sensitivity of this search is
similar to the sensitivity of the cut-based analysis and will improve
with an optimized selection. For $m_H=115$ GeV the observed (expected)
limit is 1.7 (1.2) pb with this present approach
\cite{WHmatrixD0}. Limits for other Higgs masses together with the
cut-based results are displayed in Fig.1.

\subsection{$ZH\to \ell\ell b\bar b$, $\ell=e,\mu$}

Similarly to $WH$ the Higgs boson can be produced associated with the
$Z$ boson. First we focus on the channel where the $Z$ boson decays to
a pair of electrons or muons with opposite sign. Here the $Z$ boson is
reconstructed and identified from a pair of high $p_T$ leptons with an
invariant mass constraint. Events are required to have b-tagged
jets. The dominant backgrounds result from the associated production
of a $Z$ boson with jets, among which the $Zb\bar b$ production is an
irreducible background. Other main backgrounds are $t\bar t$, $WZ$,
$ZZ$, and multijet production from QCD processes. 

In the search at \D0 \  at least two b-tagged jets are required. Cross
section limits are then derived from the dijet invariant mass
distribution within a search window. At CDF only 1 b-tagged jet is
required. After this, a two dimensional Neural Network discriminates
against the two largest backgrounds which are $Z$ + jets and $t\bar
t$. Limits are derived from the Neural Network distribution. For
$m_H=115$ GeV the observed (expected) limit is 2.7 (2.8) pb at \D0 \ 
\cite{ZHllCDF} and 2.2 (1.9) at CDF \cite{ZHllD0}, to be compared to the Standard
Model cross section expectation of 0.08 pb.

\subsection{$ZH\to \nu\nu b\bar b$, $WH\to (\ell^{\pm})\nu b\bar b$}

The $ZH\to \nu\nu b\bar b$ channel benefits from the large
$Z\to\nu\nu$ branching ratio. However it is challenging at hadron
colliders due to the absence of visible leptons and the presence of
only two jets in the final state. The two b-jets from the Higgs are
boosted along the direction of the Higgs momentum and so tend to be
more acoplanar than the dijet background. There are two major sources
of background: physics backgrounds such as $Z$+jets, $W$+jets,
electroweak diboson production or top quark production with missed
leptons and jets and the instrumental background resulting from
calorimeter mismeasurements which can lead to high $E_T$ signals with
the presence of jets from QCD processes.

A result on this search channel was presented from CDF. Selecting
events with a large $E_T > 75$ GeV and high $p_T$ b-tagged jets
(leading jet $p_T > 60$ GeV), vetoing events with isolated leptons or
where the missing $E_T$ is aligned in $\phi$ with jets eliminate much
of the physics background. Two separate analyses are optimized for one
or two b-tagged samples and later combined. Since the $WH$ channel
with an undetected lepton has the same signature those events are
taken into account in this search channel. For $m_H=115$ GeV the
expected limit at CDF is a factor 15 higher than the Standard Model
expectation \cite{ZHnnCDF}.

\begin{figure}[t]
\begin{center}
\includegraphics[width=.75\textwidth]{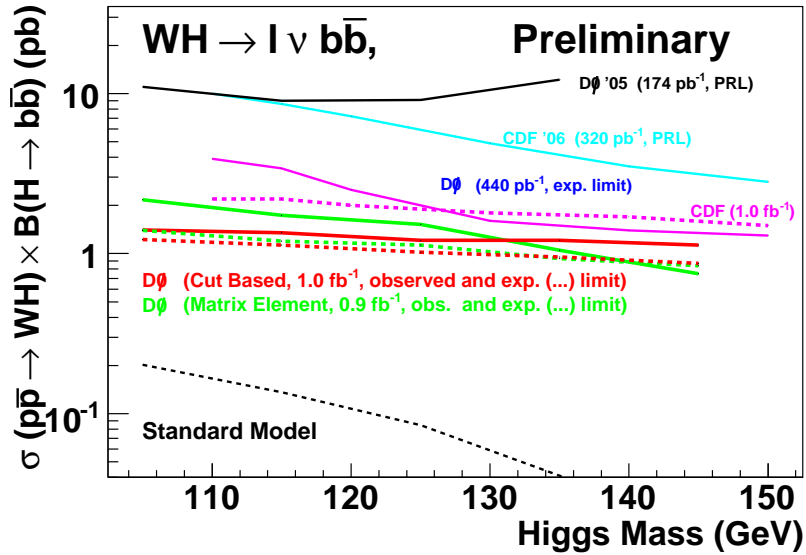}
\caption{95\% C.L. upper cross section limits on $WH$ production from the
cut-based CDF and \D0 \  measurements and from the matrix-element-approach
by \D0 .}
\end{center}  
\end{figure}

\subsection{$H\to WW^{(*)}\to \ell^+\ell^-\nu\bar\nu$, $\ell=e,\mu$}

At Higgs masses above 135 GeV the biggest branching ratio is the decay
to $WW$ pairs. With only leptons and missing energy in the final state
the main background is $WW$ production without a large overlapping QCD
background. Both, CDF and \D0 \  analyzed this channel for the three
combinations of electron and muon final states. Later the cross
section limits have been combined. 

The search strategy is to look for two high $p_T$, isolated, opposite
sign leptons, require large missing transverse energy and veto on
events with jets. Finally, the spin correlations in the decay of the
Higgs boson are used. The leptons of the Higgs decay tend to have a
small opening angle, whereas leptons from most of the backgrounds are
expected to be back-to-back. Thus a cut on the opening angle between
the leptons in the transverse plane $\Delta\phi_{\ell\ell}$ is mainly used
to discriminate against the dominant $WW$ background. Since the Higgs
mass cannot be directly reconstructed due to the neutrinos in the
final state, the cross section limit is derived from the
$\Delta\phi_{\ell\ell}$ distribution. For $m_H=160$ GeV, which yields the
best sensitivity, the expected limit at CDF \cite{HWWCDF} is a factor 6 and at \D0 \
 \cite{HWWD0,HWWmuD0} a factor of 5 higher than the Standard Model expectation.

\subsection{Combined Standard Model Higgs limits}

The above presented channels can be combined which leads to a much
more sensitive cross section limit throughout the whole discussed
mass range. Both, \D0 \ and CDF released results on the SM Higgs
combination, the obtained results can be found in \cite{comb}. A further,
important increase of the sensitivity can be gained from a combination
of the CDF and \D0 \ results. Such a first Tevatron combination limit
was released Summer 2006, the result is plotted in
Fig.2. The expected combined limits are a factor of 7.5
at $m_H=115$ GeV and a factor of 4 at $m_H=160$ GeV away from the
Standard Model expected cross sections. It should be stressed that
this result does not include CDF's new 1 fb$^{-1}$ high mass results
and it does not include any of \D0 's new 1 fb$^{-1}$ low mass
results yet. Further significant improvements are expected when all
the 1 fb$^{-1}$ results will be included. Such a new Tevatron
combination is planned for the Summer 2007.

\begin{figure}[t]
\begin{center}
\includegraphics[width=.75\textwidth]{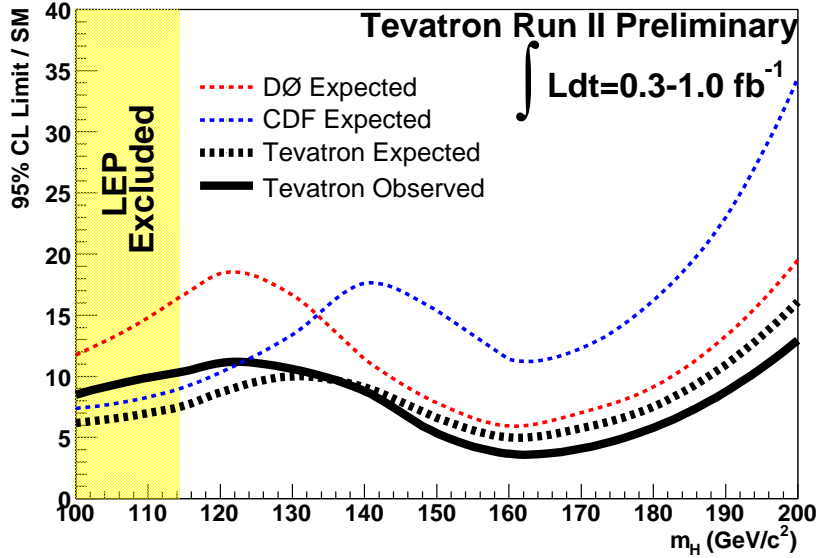}
\caption{Combined \D0 \ and CDF upper limits on Standard Model Higgs boson production.}
\end{center}  
\end{figure}

\section{MSSM Higgs searches}

The Minimal Supersymmetric Standard Model (MSSM) predicts two Higgs
doublets leading to five Higgs bosons: a pair of charged Higgs boson
($H^\pm$); two neutral CP-even Higgs bosons ($h$,$H$) and a CP-odd Higgs
boson ($A$). At tree level, the Higgs sector of the MSSM is fully
described by two parameters, which are chosen to be the mass of the
CP-odd Higgs, $m_A$, and $\tan\beta$ , the ratio of the vacuum
expectation values of the two Higgs doublets. The Higgs production
cross-section is enhanced in the region of low $m_A$ and high
$\tan\beta$ due to the enhanced Higgs coupling to down-type fermions.
This makes it possible to search in the MSSM for $\tau\tau$ final
states, which would be very challenging in the SM due to the large
irreducible background of $Z\to\tau\tau$. In the low $m_A$, high
$\tan\beta$ region of the parameter space, Tevatron searches can
therefore probe several MSSM benchmark scenarios extending the search
regions covered by LEP \cite{lepmssm}.

Both, CDF and \D0 \ performed a search for the neutral MSSM Higgs decaying to
$\tau$ pairs, where one of the $\tau$-leptons is decaying in the leptonic and the
other one in the hadronic mode. \D0 's result covers so far only the
$\mu$-channel, CDF's result is a combination of the electron and muon
channels, including $\tau_e\tau_\mu$. 

A set of Neural Networks (NN) is used at \D0 \  to discriminate $\tau$-leptons
from jets. An isolated muon is required, separated from the hadronic
$\tau$ with opposite sign. A cut on the visible $W$ mass removes most
of the remaining $W$ boson background. Further optimized NNs are used
for signal discrimination. In the cross section limit calculation the
output of the NNs for different tau types is used.

CDF uses a variable cone size algorithm for $\tau$ discrimination. An
isolated muon or electron is required, separated from the hadronic
$\tau$ with opposite sign. Most of the $W$ background is removed by a
requirement on the relative directions of the visible $\tau$ decay
products and the missing transverse energy. Cross section limits are
derived from the visible mass distribution. 

For both experiments the data is consistent with the background only
observation. Exclusion regions in the $\tan\beta$ -- $m_A$ plane can
be derived for different MSSM benchmark scenarios. Both experiments
obtained similar results \cite{tauCDF,tauD0,mark}. In the region of
$90 < m_A < 200$ GeV, $\tan\beta$ values larger than 40-60 are
excluded for the no-mixing and the $m_h^{max}$ benchmark
scenarios. Examples of such exclusion regions are shown in Fig.3. In
CDF's result the observed limits are weaker than the expectations due
to some excess of events in the data sample with a significance of
approximately 2$\sigma$.

\begin{figure}[t]
\includegraphics[width=.45\textwidth]{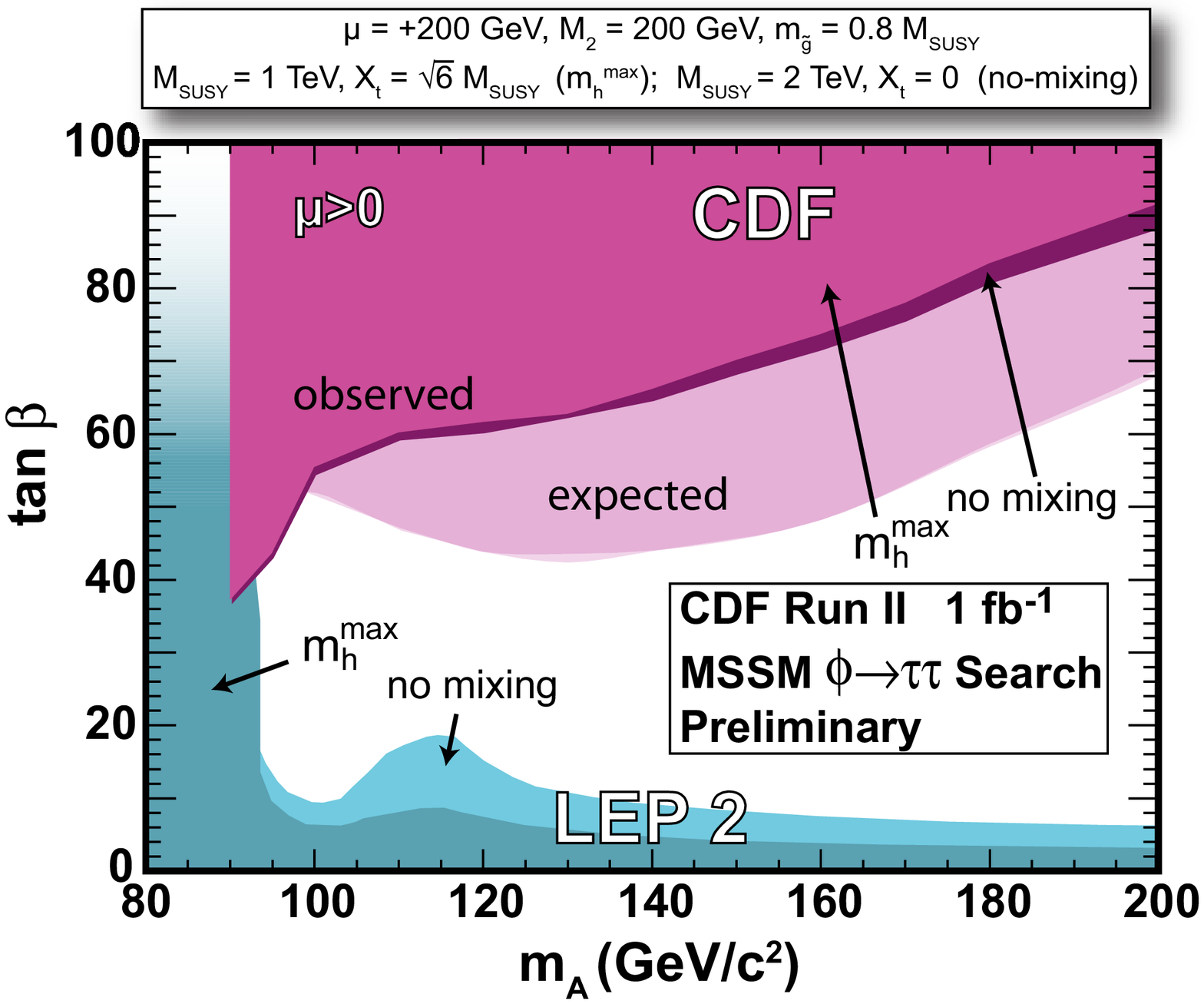}\includegraphics[width=.55\textwidth]{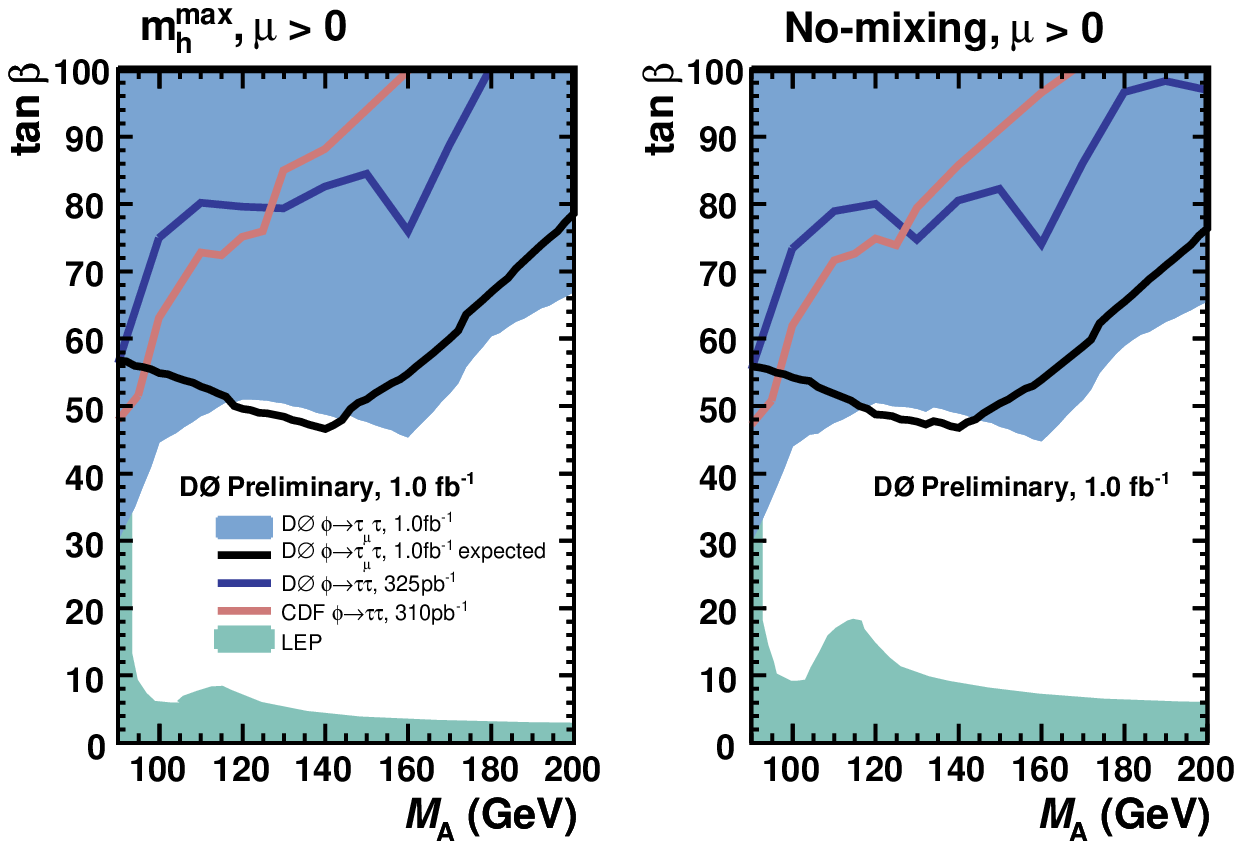}
\caption{Excluded regions by the CDF and \D0 \ experiments in the $\tan\beta$-$m_A$ plane for $\mu<0$ 
in the $m_h^{max}$ and the no-mixing scenario. }
\label{taud0} 
\end{figure}

\section{Perspectives}

Today some single channels have cross section limits similar to the
combined Tevatron results obtained half a year ago. With Tevatron's
excellent performance matching the designed delivered weekly
luminosities, a significant amount of sensitivity will be gained by an
increase of the luminosity by about a factor of 8. There is already
2.5 times more data on tape than used for the presented results. In
addition, the inclusion of more channels in the Higgs search (for
example $\tau$-final states) will gain additional sensitivity. Dijet
mass resolution, b-tagging and simulation are important ingredients
for Higgs searches and both experiments are continuously improving at
these scopes. Still a lot of improvements are expected in analyses
techniques. Especially the use of multivariate techniques, like Neural
Networks, Decision Trees and Matrix Element analyses shall bring
further important improvements. \D0 's recent evidence for Single Top
production and CDF's $WZ$ observation is an important milestone in the
use of these techniques to discriminate very low rate signals in the
presence of substantial backgrounds.

\section*{Acknowledgments}
I would like to thank my colleagues from the CDF and \D0 \ Collaborations
working on this exiting topic and for providing material for this
talk. I also like to thank the organizers of Rencontres de Moriond for
a stimulating conference and the European Union "Marie Curie"
Programme for their support.

\end{document}